\documentclass[letter,twocolumn]{jpsj3} 

\usepackage{graphicx}
\usepackage{bm}
\usepackage{color}
\usepackage{ulem}
\usepackage{amsfonts,amsthm}

\begin{document}

\newcommand{\Ha}{\mathcal{H}}
\newcommand{\mh}{\mathsf{h}}
\newcommand{\mA}{\mathsf{A}}
\newcommand{\mB}{\mathsf{B}}
\newcommand{\mC}{\mathsf{C}}
\newcommand{\mS}{\mathsf{S}}
\newcommand{\mU}{\mathsf{U}}
\newcommand{\mX}{\mathsf{X}}
\newcommand{\sP}{\mathcal{P}}
\newcommand{\sL}{\mathcal{L}}
\newcommand{\sO}{\mathcal{O}}
\newcommand{\la}{\langle}
\newcommand{\ra}{\rangle}
\newcommand{\ga}{\alpha}
\newcommand{\gb}{\beta}
\newcommand{\gc}{\gamma}
\newcommand{\gs}{\sigma}
\newcommand{\vk}{{\bm{k}}}
\newcommand{\vq}{{\bm{q}}}
\newcommand{\vR}{{\bm{R}}}
\newcommand{\vQ}{{\bm{Q}}}
\newcommand{\vga}{{\bm{\alpha}}}
\newcommand{\vgc}{{\bm{\gamma}}}
\newcommand{\bq}{\mbox{\boldmath $q$}}
\newcommand{\br}{\mbox{\boldmath $r$}}
\newcommand{\bz}{\mbox{\boldmath $z$}}
\newcommand{\bchi}{\mbox{\boldmath $\chi$}}
\newcommand{\bG}{\mbox{\boldmath $G$}}
\newcommand{\bg}{\mbox{\boldmath $g$}}

\arraycolsep=0.0em
%

\title{Ab initio Low-Dimensional Physics Opened Up by Dimensional 
Downfolding: Application to LaFeAsO 
}

\author{Kazuma \textsc{Nakamura}$^{1,2}$\thanks{Electronic mail: kazuma@solis.t.u-tokyo.ac.jp}, Yoshihide \textsc{Yoshimoto}$^{3}$, Yoshiro \textsc{Nohara}$^{4}$, Masatoshi \textsc{Imada}$^{1,2}$} 

\inst{$^{1}$ Department of Applied Physics, University of Tokyo, 7-3-1 Hongo, Bunkyo-ku, Tokyo 113-8656 \\ 
      $^{2}$ CREST, JST, 7-3-1 Hongo, Bunkyo-ku, Tokyo 113-8656 \\
      $^{3}$ Department of Applied Mathematics and Physics, Tottori University, Tottori 680-8550 \\
      $^{4}$ Department of Physics, University of Tokyo, 7-3-1 Hongo, Bunkyo-ku, Tokyo 113-0033}

\date{\today}

\abst{An {\it ab initio} downfolding method is formulated to construct low-dimensional models for correlated electrons. 
In addition to the band downfolding by constrained random 
phase approximation formulated for 3D models, 
screening away from the target layer (chain) is further involved.  
Eliminating the off-target degrees of freedom, namely, dimensional downfolding
yields {\it ab initio} low-dimensional models. 
The method is applied to derive a 2D model for a layered superconductor LaFeAsO, where the interlayer screening crucially makes the effective interaction short ranged and reduces the onsite Coulomb interactions by 10-20 \% from the 3D model for the 5 iron-$3d$ orbitals.}
\kword{%
first-principles calculation, 
effective low-dimensional Hamiltonian, 
constrained RPA method, 
dimensional downfolding, 
anisotropic low-dimensional system, 
LaFeAsO}

\maketitle
Discoveries of outstanding materials with low-dimensional (low-D) anisotropy such as cuprate \cite{Bednorz} and iron-based \cite{Hosono} superconductors have stimulated studies on low-D electronic models~\cite{Fulde-Imada}.
1D or 2D simplified models have contributed in clarifying 
strong-correlation and fluctuation effects 
characteristic of the low dimensionality.
However, to a large extent, the studies rely on {\it ad hoc} 
adjustable parameters as in the Hubbard models.

To overcome the difficulty of density functional theory (DFT) 
in strongly correlated materials, a three-stage method combining 
{\it ab initio} DFT with high-accuracy solvers for 
lattice models has been proposed~\cite{Anisimov, Aryasetiawan,Solovyev,Imai,LDA+DMFT,LDA+PIRG,Miyake1,Nakamura,Nakamura2,Miyake2,Tomczak},
stimulated by fruitful physics and potential applications 
revealed in correlated electron systems.  
In this three-stage scheme, a global band structure calculated by DFT
 such as the local density approximation (LDA) 
is downfolded into a small number of bands called ``target bands" near 
the Fermi level, for instance,  
by the constrained random phase approximation 
(cRPA)~\cite{Aryasetiawan,Solovyev,Imai,LDA+PIRG,LDA+DMFT}. 
Electronic degrees of freedom far from the Fermi level are eliminated in this step by the partial trace summation, leaving a lattice model for 
maximally localized Wannier orbitals (MLWO)~\cite{MLWF} 
for the target bands. 
This scheme has enabled quantitative studies beyond the {\it ad hoc} parameter tuning, 
when the target-band models are solved 
by high-accuracy solvers.
This {\it ab initio} downfolding has been successfully 
applied to various materials~\cite{Miyake1,Nakamura,Nakamura2,Miyake2,Tomczak}.

However, so far, the downfolding derives models in 3D space. 
It does not offer 
a real connection to the extensively studied low-D models mentioned above, thus leaving a ``missing link".

The purpose of this letter is to bridge 
the derived 3D {\it ab initio} models and low-D physics, 
thereby connect this missing link to open a new low-D studies 
from first principles. 
Our scheme is regarded as the dimensional downfolding in real space, 
supplementing the original band downfolding in energy space.
The scheme is general and works particularly well 
for quasi-low-D systems.

Advantages of 
the reduction to low-D models lie principally in numerical feasibility. 
The derived interaction for 3D models has a long-range 
tail $\propto\!1/r$,
because the conventional downfolding excludes 
metallic screenings completely. 
So, even for quasi-low-D systems, we have to consider a 3D model 
whose layers (or chains) interact with each other by long-range interactions.
The present dimensional downfolding largely simplifies the problem not only to tractable low-D models 
but also to short-range-interaction models
when metallic screening from layers (chains) 
other than the target one is considered.

As an example, we apply it to derive an effective 2D model
 for LaFeAsO.
We show that interlayer screenings reduce onsite 
Coulomb interactions by 10-20 \% 
and further remove the long-range part of the screened interaction.
Our formalism justifies a multi-band 2D Hubbard model for LaFeAsO 
from first principles.

Below we take an example of quasi-2D systems.
Our goal is to derive an effective low-energy 2D model as
\begin{eqnarray}
&&\mathcal{H}
= \sum_{\sigma} \sum_{{\bf R} {\bf R'}} \sum_{nm}  
  t_{m {\bf R} n {\bf R}'} 
                   a_{n {\bf R}}^{\sigma \dagger} 
                   a_{m {\bf R'}}^{\sigma}   \nonumber \\
&&+ \frac{1}{2} \sum_{\sigma \rho} 
  \sum_{{\bf R} {\bf R'}} \sum_{nm} 
  \biggl\{ U_{m {\bf R} n {\bf R}'} 
                   a_{n {\bf R}}^{\sigma \dagger} 
                   a_{m {\bf R'}}^{\rho \dagger}
                   a_{m {\bf R}'}^{\rho} 
                   a_{n {\bf R}}^{\sigma} \nonumber \\ 
&&+J_{m {\bf R} n {\bf R}'} 
\bigl(\!a_{n {\bf R}}^{\sigma \dagger} 
      \!a_{m {\bf R'}}^{\rho \dagger}
      \!a_{n {\bf R}}^{\rho} 
      \!a_{m {\bf R}'}^{\sigma} 
  \!+\!a_{n {\bf R}}^{\sigma \dagger} 
     \!a_{n {\bf R}}^{\rho \dagger}
     \!a_{m {\bf R}'}^{\rho} 
     \!a_{m {\bf R}'}^{\sigma}\bigr)\! \biggr\}, 
\label{eq:H}                
\end{eqnarray}
where $a_{n {\bf R}}^{\sigma \dagger}$ ($a_{n {\bf R}}^{\sigma}$) 
is a creation (annihilation) operator of a target-band electron with 
spin $\sigma$ at the $n$th MLWO in the unit cell at ${\bf R}$ in the 2D layer.
The single-particle levels and transfers are given 
by $t_{m {\bf R} n {\bf R}'}$.
Screened Coulomb and exchange interactions are 
$U_{m {\bf R} n {\bf R'} }$=$\langle 
    \phi_{m {\bf R}}  \phi_{n {\bf R}'} | W | 
    \phi_{m {\bf R}}  \phi_{n {\bf R}'} \rangle$
and 
$J_{m {\bf R} n {\bf R}'}$=$\langle \phi_{m {\bf R}} \phi_{n {\bf R}'} | W | \phi_{n {\bf R}'} 
\phi_{m {\bf R}} \rangle$,
respectively, with $| \phi_{n {\bf R}} \rangle$
=$a_{n {\bf R}}^{\dagger}|0\rangle$, 
where $W$ is an effective electron Coulomb interaction for the 2D model. 

We employ cRPA in a two-step procedure to derive $W$ for the 2D model. 
First, we follow the conventional band downfolding~\cite{Aryasetiawan} to derive the 3D model for the target bands: We divide the total polarization into two parts; a polarization $\chi^t$ within the target bands and the rest $\chi^r$. 
The division is well-defined when the target bands 
are isolated from other bands.
In the usual cRPA for deriving a 3D model (3D-cRPA), 
$\chi^t$ is excluded and the screened interaction is 
calculated from $W_r=v/(1-\chi^r v)$ with the bare Coulomb interaction $v$.  
\begin{figure}[tbh]
	\begin{center}
 	\includegraphics[width=0.37\textwidth]{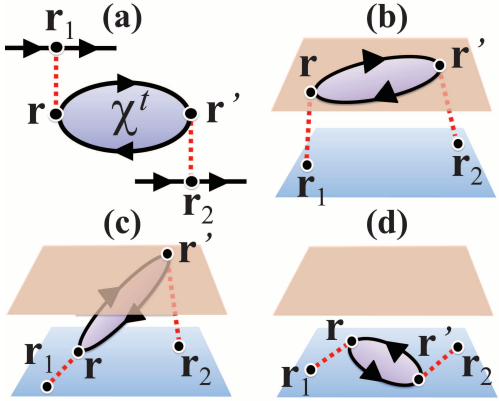}  
 	\end{center}
\caption{(Color online)  
Schematic diagram for effective interactions between electrons at 
${\bf r}_1$ and ${\bf r}_2$ in the target layer, 
screened by intra- and inter-layer polarizations $\chi^t({\bf r}, {\bf r}')$.
Dotted lines are a bare Coulomb interaction, 
while solid lines with arrows are electron propagators. 
See the text for details.
}
\label{fig:Fig1}
\end{figure}

In the next cRPA step to derive a 2D model for the target layer (2D-cRPA), 
we eliminate degrees of freedom in the other layers 
after including the screening by the electrons in the other layers.  
Figure~\ref{fig:Fig1}~(a) shows an RPA diagram (symbolically for the second-order one) of the effective interaction between electrons at ${\bf r}_1$ and ${\bf r}_2$ in the target layer while ${\bf r}$ and ${\bf r}'$ denote 
polarization points of $\chi^t$.
The electrons at ${\bf r}_1$ and ${\bf r}$ (or ${\bf r}'$ and ${\bf r}_2$)
 interact via $v$ (dotted line).  
The screening of the intralayer interaction within the target layer is classified into several processes; 
the panel~(b) shows the screening by a polarization in the other layers, while the panel~(c) describes that by an interlayer polarization between the target and the other layers. 
The panel~(d) is the screening 
by a polarization within the target layer itself.
This process 
should be excluded in the present 2D-cRPA downfolding, while (b) and (c) are included.  

In the first step, the 3D-cRPA becomes accurate and the vertex correction can be ignored, when the eliminated band energy in the denominator of the propagator is far from the Fermi level. In the second step, the 2D-cRPA becomes accurate if the numerators are small 
in the vertex corrections derived from small interlayer components of
renormalized transfers and/or interactions. The small parameter in the vertex correction is roughly the ratio of the screened interlayer to onsite interactions obtained by full RPA. We do not consider it here.

\begin{figure*}[htb]
	\begin{center}
	\includegraphics[width=1.0\textwidth]{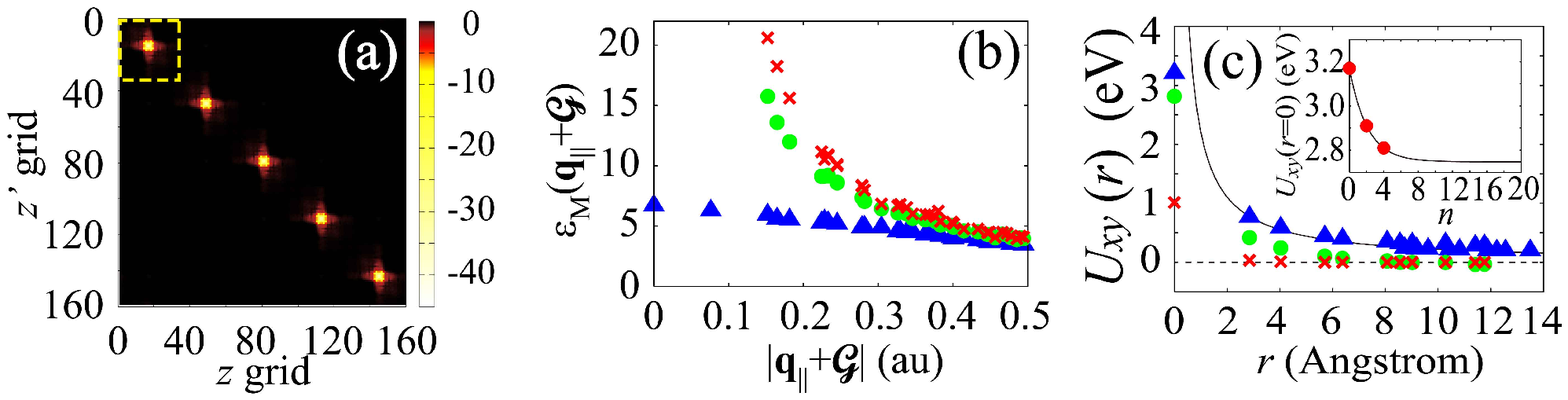}  

	\end{center}
\caption{(Color online) 
(a) $\chi^t_{{\bf G}_{\|} {\bf G}_{\|} {\bf q}_{\|}}(z,z')$ 
(in a.u.) in the $z$-$z'$ plane with 
${\bf G}_{\|}$=${\bf G}_{\|}$=${\bf q}_{\|}$={\bf 0}. 
Inside the dashed line is the cutting region defined in eq.~(\ref{eq:cut}). 
(b) Dielectric functions as a function of $|{\bf q_{\|}}+{\cal G}|$ obtained by full-RPA [(red) crosses], 2D-cRPA [(green) dots], and 3D-cRPA [(blue) triangles]. (c) Screened Coulomb interactions of LaFeAsO as a function of distance $r$ between centers of MLWOs. Only the interactions between $d_{xy}$ MLWOs, $U_{xy}(r)$$\equiv$$U_{xy{\bf 0}xy{\bf R}}$, are plotted.  The intersite interactions are nearly the same between the inter- and intra-orbital pairs. 
The same symbols as (b) are used for interaction values. The 3D-cRPA interaction is scaled as $1/\bigl[ \epsilon_{{\rm M}}^{{\rm 3DcRPA}}({\bf q}\!\to\!0) r \bigr]$ (solid curve). (Inset) Convergence of onsite intraorbital $U_{xy}(r\!=\!0)$ as a function of number of screening layers $n$ to $n\rightarrow \infty$. Solid curve is the fitted $U(n)$ (see the text).}
\label{fig:Fig2}
\end{figure*}

Practical calculations proceed as follows: 
The target-band polarization $\chi^t$ is given by 
\begin{eqnarray}
\chi^t_{{\bf G} {\bf G}'}({\bf q})&=&
{\sum_{{\bf k}} \sum_{\alpha \beta}}^\prime 
\langle \psi_{\alpha {\bf k}+{\bf q}} 
| e^{i ({\bf q}+{\bf G}) {\bf r}} 
| \psi_{\beta {\bf k}} \rangle \nonumber \\
&\times&
\langle \psi_{\beta {\bf k}} 
| e^{-i ({\bf q}+{\bf G'}) {\bf r}} | 
\psi_{\alpha {\bf k}+{\bf q}} \rangle 
\frac{ f_{\alpha {\bf k}+{\bf q}} - f_{\beta {\bf k}}} 
{E_{\alpha {\bf k}+{\bf q}} - E_{\beta {\bf k}}},
\label{eq:chi} 
\end{eqnarray}
where $\{\psi_{\alpha {\bf k}} \}$, $\{ E_{\alpha {\bf k}} \}$, and 
 $\{ f_{\alpha {\bf k}} \}$ are the Bloch states, their energies, and 
 occupancies, respectively,  
and the prime summation runs over the target bands only.
The idea for the dimensional downfolding is to use the Fourier transform along the interlayer $c$ axis ($z$ grids) representing the axis complementary to the target layer~\cite{1Dcase},   
\begin{eqnarray}
\!\chi^t_{{\bf G_{\|}}\!{\bf G'_{\|}}\!{\bf q_{\|}}}\!(z,\!z')\!
\!=\!\sum_{\!{q_{\perp}}\!{{\rm G}_{\perp}}\!{{\rm G}'_{\perp}}}\! 
\!e^{i ({\rm q}_{\perp}\!+\!{\rm G}_{\perp}\!) z} 
 \chi^t_{{\bf G}\!{\bf G'}}\!({\bf q}) 
e^{-i ({\rm q}_{\perp}\!+\!{\rm G}'_{\perp}\!) z'\!} 
\label{eq:chi_z}
\end{eqnarray} 
with ${\bf q}={\bf q}_{\|}+{\bf q}_{\perp}$ and ${\bf G}={\bf G}_{\|}+{\bf G}_{\perp}$. Here, $\|$ ($\perp$) represents an orientation parallel 
(complementary) to the layer. 
Suppose the interlayer hopping is switched off. Then
$\chi^t_{{\bf G_{\|}} {\bf G'_{\|}} {\bf q_{\|}}} (z, z')$ is block diagonal in the $z$-$z'$ plane. In quasi-2D materials, 
this block diagonal structure 
is still essentially retained as we see in Fig.~\ref{fig:Fig2},
which enables excluding the polarization within the target layer as   
\begin{eqnarray}
\!\tilde{\chi}^t_{{\bf G_{\|}}\!{\bf G'_{\|}}\!{\bf q_{\|}}}(z,\!z')\! 
=\chi^t_{{\bf G_{\|}}\!{\bf G'_{\|}}\!{\bf q_{\|}}}\!(z,\!z')
\Lambda_{{\rm cut}}(z,\!z')
\label{eq:chi_tild_z}
\end{eqnarray}
with 
\begin{eqnarray}
\Lambda_{{\rm cut}}(z,z')
  = \left\{
    \begin{array}{@{\,}ll}
      \mbox{%
        \parbox{0.5cm}{%
          $\displaystyle
          0
          $
        }
      } & \mbox{($z,z'\in$ target layer),}
    \\[+5pt]
      \displaystyle 1
      & \mbox{(otherwise).} 
    \end{array}
  \right.
\label{eq:cut}
\end{eqnarray}
The target layer is 
defined by a unit layer complementary to the $c$ axis, 
bounded by the middle 
of the Bravais-lattice points.
Then, we go back to the reciprocal space with the inverse Fourier transform 
\begin{eqnarray}
\!\tilde{\chi}^t_{{\bf G_{\|}}\!{\bf G'_{\|}}\!{\bf q_{\|}}}\!
 (g_{\perp},g'_{\perp})\!
\!=\!\sum_{z,z'}\! 
e^{-i g_{\perp} z} 
 \tilde{\chi}^t_{{\bf G_{\|}}\!{\bf G'_{\|}}\!{\bf q}_{\perp}}\!(z,z') 
e^{i g'_{\perp} z'}\!
\label{eq:chi_tild_g}.
\end{eqnarray} 
Since the original lattice periodicity along the 
$c$ axis is broken due to the cutting, 
in this transform, reciprocal lattice vector $\bg_{\perp}$ for the superlattice is used, 
instead of ${\bf q_{\perp}}$+${\bf G_{\perp}}$. 
The symmetric dielectric function~\cite{Louie} is 
\begin{eqnarray}
\tilde{\epsilon}_{{\cal G} {\cal G'}}(\!{\bf q}_{\|}\!) 
\!=\!\delta_{{\cal G}{\cal G'}} 
\!-\! \frac{ \tilde{\chi}^t_{{\bf G_{\|}} {\bf G'_{\|}} {\bf q_{\|}}} 
(g_{\perp},\!g'_{\perp}\!)\!+\!
 \chi^{r}_{{\bf G_{\|}} {\bf G'_{\|}} {\bf q_{\|}}} 
(g_{\perp},\!g'_{\perp}\!)}{|{\bf q_{\|}}+{\cal G}||{\bf q_{\|}}+{\cal G'}|}
\label{eq:epsilon}
\end{eqnarray}
with $\cal{G}\!=\!{\bf G_{\|}}\!+\!{\bg_{\perp}}$ and $\chi^r$ being the 3D-cRPA polarization. 
Matrix elements of the screened Coulomb interaction are 
\begin{eqnarray}
    U_{m {\bf 0} n {\bf R}} 
\!=\!\frac{4 \pi}{V}\!\sum_{{\bf q_{\|}} {\cal G} {\cal G'}}\! 
 e^{-i {\bf q_{\|}} {\bf R}} 
\rho_{m {\bf q_{\|}}}({\cal G}) 
\tilde{\epsilon}_{{\cal G}{\cal G'}}^{-1}(\!{\bf q_{\|}}\!) 
\rho_{n {\bf q_{\|}}}^{*}({\cal G'}\!),
\label{eq:matrix_U}
\end{eqnarray}
where $V$ is a crystal volume and 
\begin{eqnarray*}
\rho_{m {\bf q_{\|}}}({\cal G})\!=\! 
\frac{1}{N_{\|}} 
\sum_{{\bf k_{\|}}}^{N_{\|}} 
\frac{\langle \tilde{\psi}_{m {\bf k_{\|}}\!+\!{\bf q_{\|}}\!} 
| e^{i ({\bf q_{\|}}+{\cal G}) {\bf r}} 
| \tilde{\psi}_{m {\bf k_{\|}}} \rangle}
{|{\bf q_{\|}} + {\cal G}|}  
\end{eqnarray*}
with  
$
|\tilde{\psi}_{n {\bf k_{\|}}}\rangle = 
\sum_{{\bf R}}^{N_{\|}} 
|\phi_{n {\bf R}}\rangle e^{-i {\bf k_{\|} R}}$. 
The exchange term is
\begin{eqnarray}
J_{m {\bf 0} n {\bf R}} 
\!=\!\frac{4 \pi}{V} \sum_{{\bf q_{\|}} {\cal G} {\cal G'}}
\rho_{mn\!{\bf R}{\bf q_{\|}}}({\cal G}) 
\tilde{\epsilon}_{{\cal G} {\cal G'}}^{-1}({\bf q_{\|}}) 
\rho_{mn\!{\bf R}{\bf q_{\|}}}^{*}\!({\cal G'}) 
\label{eq:matrix_J}
\end{eqnarray}
with 
\begin{eqnarray*}
\rho_{mn {\bf R} {\bf q_{\|}}} ({\cal G}) \!=\! \sum_{{\bf k_{\|}}}^{N_{\|}} e^{-i {\bf k_{\|}} {\bf R}} \frac{\langle \tilde{\psi}_{m\!{\bf k_{\|}}\!+\!{\bf q_{\|}}} | e^{i ({\bf q_{\|}}\!+\!{\cal G}){\bf r}} | \tilde{\psi}_{n\!{\bf k_{\|}}} \rangle}{N_{\|} |{\bf q_{\|}}\!+\!{\cal G}|}. 
\end{eqnarray*}

The self-energy due to off-target propagators may also be calculated in 2D-cRPA, which modifies the dispersion.  However, in the same spirit with neglecting the vertex correction, we do not consider it.  
On the other hand, to avoid the double counting, the interaction already considered in LDA has to be subtracted, where it is dominantly for 
the Hartree term and can easily be implemented~\cite{Misawa}.

Now we apply this formalism to derive a 2D model for LaFeAsO. {\it Ab initio} 
density-functional~\cite{DFT} calculations were performed with plane-wave-basis-set code of {\it Tokyo Ab initio Program Package}~\cite{TAPP} with LDA~\cite{PW92} and norm-conserving pseudopotentials.~\cite{PP1} 
The details for an iron pseudopotential are 
found in ref.~\citen{Miyake2}.  
The experimental structure of LaFeAsO was taken from ref.~\citen{LaFeAsO}.
The cutoff energies in wavefunctions and charge densities were set to 
 100 Ry and 900 Ry, respectively, and a 5$\times$5$\times$5 
 $k$-point sampling was employed.   
The polarization was expanded in plane waves with a 20-Ry cutoff and the total number of bands considered in the polarization was set to 130. 
The Brillouin-zone (BZ) integral on wavevector was evaluated by 
 the generalized tetrahedron method~\cite{Fujiwara}. 

We remark some details. 
Sufficient cutoff energy is necessary for correct decay of $\chi^t_{{\bf G} {\bf G'}}({\bf q})$ at large ${\bf G}-{\bf G'}$. Otherwise a noise in its Fourier transform $\chi^t_{{\bf G_{\|}}\!{\bf G'_{\|}}\!{\bf q_{\|}}}\!(z,\!z')$ of eq.~(\ref{eq:chi_z}) brings about an incorrect depolarization in the inverse transform $\tilde{\chi}^t_{{\bf G_{\|}}\!{\bf G'_{\|}}\!{\bf q_{\|}}}\! (g_{\perp},g'_{\perp})\!$, obtained after the cutting [Eq.~(\ref{eq:chi_tild_z})]. We confirmed that a cutoff as large as 20 Ry is sufficient in the present case. We also need a careful treatment of poles at $E_{\alpha {\bf k+q}}\!=\!E_{\beta {\bf k}}$ in eq.~(\ref{eq:chi}), for which, to reduce an error by discretized $k$ sum, we utilize
\begin{eqnarray}
\frac{ f_{\alpha {\bf k}+{\bf q}} - f_{\beta {\bf k}}} 
{E_{\alpha {\bf k}+{\bf q}} - E_{\beta {\bf k}}}
\sim \frac{d f(E)}{d E} \Bigr|_{E = E'}  
= \delta \bigl( E' - E_{{\rm F}} \bigr)
\end{eqnarray}
with $E'$=$(E_{\alpha {\bf k}+{\bf q}}\!+\!E_{\beta {\bf k}})/2$, based on the central-difference approximation to $df(E)/dE$ at $E = E'$.

Figure~\ref{fig:Fig2}~(a) shows a plot of 
$\chi^t_{{\bf G}_{\|} {\bf G}_{\|} {\bf q}_{\|}}(z,z')$
in eq.~(\ref{eq:chi_z}) with 
${\bf G}_{\|}$=${\bf G}_{\|}$=${\bf q}_{\|}$={\bf 0}. 
We see five bright spots, where polarizations are large. 
The matrix is nearly block diagonal, 
indicating that the target-band polarization is 
confined in each layer.
Thus, we can successfully exclude the polarization 
formed within the target layer.  
The cutting region defined in eq.~(\ref{eq:cut}) 
is that inside the dashed line. 

We plot in Fig.~\ref{fig:Fig2}~(b) diagonal elements of 2D-cRPA 
macroscopic dielectric function 
$\epsilon_{{\rm M}} = 1/\tilde{\epsilon}^{-1}$ [(green) dots] 
obtained via inverse of eq.~(\ref{eq:epsilon}).
We also show a full-RPA result [(red) crosses] 
obtained without cutting and a 3D-cRPA result [(blue) triangles]
 obtained by totally excluding $\chi^t$ from the screening. 
We see diverging $\epsilon_{{\rm M}}$ in the ${\bf q_{\|}}$+${\cal G}$$\to$0 limit in the full-RPA and 2D-cRPA results, 
indicating the metallic screening in both of the cases.

Figure~\ref{fig:Fig2}~(c) displays matrix elements of a 
2D-cRPA screened Coulomb interaction, 
$U_{m {\bf 0} n {\bf R}}$ in eq. (\ref{eq:matrix_U}), 
denoted by (green) dots, for LaFeAsO as 
a function of the distance between the centers of the MLWOs; 
$r=|\langle \phi_{n {\bf R}}| \br | \phi_{n {\bf R}}\rangle
- \langle \phi_{m {\bf 0}}| \br | \phi_{m {\bf 0}} \rangle |$. 
In this plot, we exemplify the case $m$=$n$=$d_{xy}$. 
Intersite interactions were found to be orbital independent.
Full-RPA [(red) crosses] and 3D-cRPA [(blue) triangles] 
results are also plotted.

The interlayer screening generates a qualitatively different feature in the model: It turns the long-ranged interaction in 3D-cRPA into short ranged. We find that for LaFeAsO the interactions larger than 0.1 eV are limited in the region up to the next-nearest neighbors ($r \le 4 \AA$), implying the screening length $\sim 4 \AA$ determined by the interlayer channel.  
It justifies the use of a 2D short-ranged-interaction model from first principles.

A careful analysis is needed to reach the convergence for the number of screening layers $n$. 
Taking five $k_{\perp}$ points in BZ, for instance,
simulates the number of screening layers $n=2$
each above and below the target layer in a supercell.
Since we need the interaction parameter in the limit $n\rightarrow \infty$,
the data is extrapolated by $U(n)=U(\infty)-[U(\infty)-U(0)]e^{-(n/\sigma)}$.
Here, $\sigma$ represents the effective screening length in the layer unit and
$U(0)$ is the 3D-cRPA value.
The inset of Fig.~\ref{fig:Fig2}~(c) shows the fitting of the onsite intraorbital $U_{xy}$.
We obtain $U(\infty)$ = 2.75 eV and $\sigma$ = 2.1.
The interlayer screening reduces the onsite $U$
by 10 to 20 \% from the 3D-cRPA values. 

\begin{table*}[htb]
\caption{Comparison between onsite 3D-cRPA and 2D-cRPA interaction parameters for eq.~(\ref{eq:H}). Upper (lower) two rows give diagonal 
(off-diagonal) interaction. 
The unit is eV.  In the column labels, 
$(1,2,3,4,5) := (xy,yz,3z^2-r^2,zx,x^2-y^2)$.} 
\label{PARAM}
\centering 
\begin{tabular}{ccccccccccccccccc}
\hline \hline 
\\ [-3mm]
 & 11 & 22 & 33 & 44 & 55 & 12 & 23 & 34 & 45 & 13 & 24 & 35 & 14 & 25 & 15 \\
 \hline
 $U_{{\rm 3D}}$ &\ 3.17& 2.66& 3.14& 2.66& 2.10& 1.95& 2.14& 2.14& 1.67& 1.94& 1.79& 1.68& 1.95& 1.67& 1.99\\
 $U_{{\rm 2D}}$ &\ 2.75& 2.26& 2.74& 2.24& 1.68& 1.55& 1.74& 1.74& 1.27& 1.52& 1.37& 1.28& 1.55& 1.27& 1.59\\
 $J_{{\rm 3D}}$ &\  -\ \ &\  -\ \ &\  -\ \ &\  -\ \ &\  -\ \ & 0.48 & 0.38   & 0.38  & 0.38  & 0.56  & 0.40  & 0.43  & 0.48  & 0.38  & 0.26  \\
 $J_{{\rm 2D}}$ &\  -\ \ &\  -\ \ &\  -\ \ &\  -\ \ &\  -\ \ & 0.48  & 0.38  & 0.38  & 0.38  &  0.56 & 0.40  & 0.43  & 0.48  & 0.38  & 0.26  \\
\hline \hline
\end{tabular}
\end{table*}
Table~\ref{PARAM} summarizes the onsite interaction parameters 
derived from 3D-cRPA and extrapolated 2D-cRPA.  
The diagonal interactions are compared in the upper two rows.
The interlayer screening is nearly orbital independent and $U_{2D}$ is uniformly reduced by $\sim$0.4 eV from $U_{{\rm 3D}}$.
This is because the screening length is much larger than 
the spread of MLWOs.  
In contrast to the diagonal part of the interaction, 
the interlayer screening does not affect 
off-diagonal exchange interactions (lower two rows).

Recent variational Monte Carlo simulation \cite{Misawa} has revealed 
that the use of the 3D-cRPA parameters \cite{Nakamura} results 
in an overestimate of the antiferromagnetic ordered moment as $\sim$2.0 $\mu_{{\rm B}}$ in contrast to the experimental value 0.3-0.6 $\mu_{{\rm B}}$ \cite{LaFeAsO,Qureshi}.
A part of the discrepancy is solved by considering La-4$f$-electron screening ignored in the 3D-cRPA calculation 
in ref.~\citen{Nakamura}, yielding 10\% reduction 
of $U$ \cite{Miyake2}. 
However, the present 2D-cRPA is crucial in further reducing
10-20 \% of the $U$ values, 
thus leading to a nearly perfect agreement with the experiment.
An application to $\kappa$-ET conductors also indicating that this method offers a promising versatile scheme will be discussed elsewhere. 
The present method opens up a way to study low-dimensional physics from first principles. 


In summary, we have formulated a general {\it ab initio} scheme to derive an effective low-D model by a downfolding of a 3D model. 
The formalism eliminates the degrees of freedom for layers (chains) other than the target one, leaving an {\it ab initio} low-D model.
After decomposing the polarization in the target band into 
layer/chain-by-layer/chain contributions in the real space,
the low-D model is obtained with the interaction screened by the interlayer/chain channel from RPA polarizations away from the target layer/chain. 
For metallic systems such as LaFeAsO, this screening is essential because it deletes the long-range part of the interactions.
These results justify the 2D short-ranged models as effective {\it ab initio} models.
This scheme enables first-principles studies of strongly correlated materials with low-D anisotropies heretofore extensively 
studied only by models with {\it ad hoc} parameters.

\acknowledgements{
This work is supported from MEXT Japan under the grant numbers 22740215. 
All the computations have been performed on Hitachi SR11000 at the Supercomputer Center of ISSP, University of Tokyo. We thank Ryotaro Arita, Hiroshi Shinaoka, and Takahiro Misawa for useful discussions.}




\begin{thebibliography}{9}
\bibitem{1Dcase} 
In the 1D case, the Fourier transform is performed for the 2D plane complementary to the target chain.
\end{thebibliography}
\end{document}